\documentclass[%
 reprint,
superscriptaddress,
 amsmath,amssymb,
 aps,
floatfix,
aip,jcp,
]{revtex4-1}

\usepackage{graphicx}
\usepackage{dcolumn}
\usepackage{bm}
\usepackage{xfrac}
\usepackage{gensymb}
\usepackage{multirow}
\usepackage{booktabs}
\usepackage{float}
\usepackage{makecell}
\usepackage{siunitx}
\usepackage{threeparttable}
\usepackage{enumerate}
\usepackage[final]{changes}

\usepackage{python}

\newcommand{\CASTEP}{CASTEP}

\newcommand{\etal}{\emph{et al} }
\newcommand{\mystrut}{\rule{0pt}{3.0ex}\rule[-1.2ex]{0pt}{0pt}}
\newcommand{\myline}{\Xhline{0.25\arrayrulewidth}}

\begin{document}
\title{Regularized SCAN functional}

\author{Albert P. Bart\'ok}
\affiliation{%
Scientific Computing Department\\
Science and Technology Facilities Council, Rutherford Appleton Laboratory, Didcot, OX11 0QX, United Kingdom}
 \email{apbartok@gmail.com}
\author{Jonathan R. Yates}%
\affiliation{%
Department of Materials\\
University of Oxford, Oxford OX1 3PH, United Kingdom}%

\date{\today}

\begin{abstract}
We propose modifications to the functional form of the SCAN density functional to eliminate numerical instabilities. This is necessary to allow reliable, automatic generation of pseudopotentials (including PAW potentials). The regularized SCAN is designed to match the original form very closely, and we show that its performance remains comparable.
\end{abstract}

                              \maketitle


\section{\label{sec:intro}Introduction}
First principles modelling of electronic structure has become a standard tool in studying the structure, stability and dynamics of matter on the atomistic scale, with Density Functional Theory (DFT) being particularly popular, due to the balance of computational accuracy and cost\cite{Jones:2015gg}. The major source of inaccuracy in Kohn-Sham DFT calculations\cite{Kohn:1965uia} is the necessity of using the exchange-correlation functional, which for general systems, only exists in approximate forms. Semilocal functionals based on the Generalized Gradient Approximation (GGA), for example the Perdew-Burke-Ernzerhof (PBE) functional\cite{Perdew:1996gs}, model the electronic structure at a reasonable accuracy for a wide range of problems. However, there is a need for functionals with yet greater accuracy. Compared to GGAs, the meta-generalized Gradient Approximations (mGGA) provide more flexibility in the approximate functional form by introducing another local property on which the exchange-correlation functional depends, the orbital kinetic energy density, in addition to the electron density and its gradients. The recently proposed Strongly Constrained and Appropriately Normed (SCAN) functional is the first mGGA constructed such that \added{it satisfies} all known constraints \replaced{that a semi-local functional can satisfy}{on model systems are satisfied}, and the remaining free parameters are fitted to reproduce exact or accurate reference values, or norms, of exchange and correlation energies. The resulting functional has proved broadly transferable\cite{Sun:2016jp}, and improved the DFT description of a wide range of systems, such as liquid water and ice,\cite{Chen:2017jn} semiconductor materials\cite{Remsing:2017fy} or metal oxides\cite{Gautam:2018gx}.

Despite the tremendous success of SCAN, its implementation in DFT packages intended for condensed matter simulations is, at the time of writing this manuscript, somewhat limited.
For example, most recent versions of the all-electron general potential linearized augmented plane wave (LAPW) codes \verb+elk+\cite{elk} and \verb+WIEN2K+\cite{wien2k} only allow non-selfconsistent calculations with mGGA functionals. 
To date, in plane-wave pseudopotential DFT implementations the availability of SCAN-based pseudopotentials has also been limited, and to our knowledge, only a norm-conserving library exists\cite{Yao:2017gn}, which is lacking kinetic energy density augmentation terms and non-linear core corrections.
For this reason, many calculations published on condensed phase simulations use PBE pseudopotentials\cite{Fu:2018cy,Chen:2017jn,Janthon:2014dr,Dorner:2018cp,Bonati:2018iu,Isaacs:2018hm}, which at best is an uncontrolled approximation.
This type of inconsistency in using pseudopotentials has been studied by Fuchs \etal\cite{Fuchs:1998cf}, and they have shown that using LDA pseudopotentials in GGA calculations leads to significant errors in the calculated structural properties.
We have also found earlier\cite{bartok2019ultrasoft} that all-electron properties are much more accurately reproduced when consistent pseudopotentials are used.

Our motivation for this current work was to generate a library of SCAN ultrasoft pseudopotentials for the entire periodic table, based on our previous work \cite{bartok2019ultrasoft}.
However, we found severe numerical instabilities in both the solution of the atomic all-electron generalized Kohn-Sham equation, which is normally the first step in the pseudopotential generation workflow, and again during the pseudopotential construction itself. Indeed, it has been previously observed that SCAN is numerically less stable than GGA exchange-correlation functionals\cite{Yang:2016ef}, and recent work has identified shortcomings of the iso-orbital indicator component of some mGGA functionals\cite{Furness:2019ic}.
To remedy this situation, we propose a regularized form of the original SCAN functional (rSCAN), which retains the accuracy of the original form, while improving its stability.

In this paper we analyze the properties of the iso-orbital indicator of SCAN, used to connect different approximations of the exchange-correlation energy based on the local bonding environment.
We describe a modification which eliminates the unphysical divergence of the exchange-correlation potential which occurs in some free atoms, while keeping the iso-orbital indicator close to the original expression for most regions.
We also identify a feature of the switching function in SCAN which introduces rapidly oscillating regions in the exchange-correlation potential.
This causes instabilities in the pseudopotential generation procedure and also affects the discrete representation of the potential on a Fourier grid, which is pivotal in DFT programs using a plane-wave basis set.
We propose a small modification, which provides smoother switching, while retaining the superb description of the exchange-correlation energy of the original SCAN functional.
We test the rSCAN to establish its closeness to the original form, and provide benchmark calculations of fully consistent plane-wave pseudopotential DFT with rSCAN.

\section{Regularized SCAN}
\subsection{The iso-orbital indicator function}
A crucial ingredient in SCAN\cite{Sun:2015ef} and some other mGGA functionals\cite{Sun:2013ku} is the iso-orbital indicator function, defined as
$\alpha=\frac{\tau-\tau_W}{\tau_U}$,
with the definitions of the used quantities listed in Table~\ref{tab:defs}.
\begin{table}[tbh]
\begin{ruledtabular}
\begin{tabular}{c|c}
Kohn-Sham orbitals & $\psi_i$ \mystrut\\
\myline
orbital kinetic energy density & $\tau =  \sfrac{1}{2} \sum_i^\textrm{occ} |\nabla \psi_i|^2$
\mystrut\\
\myline
electron density & $n=\sum_i ^\textrm{occ} |\psi_i|^2$ 
\mystrut\\
\myline
Weizs\"acker kinetic energy density & $\tau_W = \frac{|\nabla n|^2}{8 n}$
\mystrut\\
\myline
\makecell{kinetic energy density of \\the uniform electron gas} &
$\tau_U=(\sfrac{3}{10})(3\pi^2)^{\sfrac{2}{3}}n^{\sfrac{5}{3}}$\mystrut
\end{tabular}
\end{ruledtabular}
\caption{Definition of the quantities based on the Kohn-Sham orbitals used in this work.}
\label{tab:defs}
\end{table}
$\alpha$ detects different local bonding environments, such as covalent single, metallic or weak bonds, and is used to switch between various local approximations of the exchange and correlation energies, derived for the appropriate bonding type.

\begin{figure}[ht]
    \includegraphics[width=\linewidth]{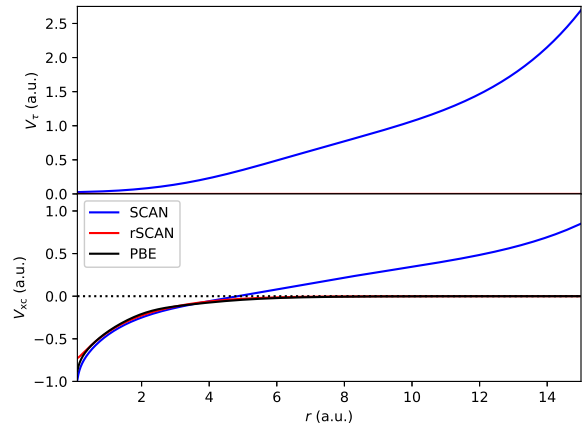}
    \caption{SCAN and rSCAN exchange-correlation potentials computed on densities corresponding to a \replaced{singly occupied}{single} $1s$ orbital. PBE is also shown for reference.}
    \label{fig:1s}
\end{figure}

However, Furness and Sun\cite{Furness:2019ic} have found that derivatives of $\alpha$ with respect to the electron density display divergent behaviour at rapidly decreasing electron densities, such as in some free atoms at large distance from the nucleus.
The consequence is that the potential itself becomes divergent in these cases, as the derivatives of $\alpha$ appear in the expressions for the potential, and are not dampened sufficiently by other terms. Therefore the resulting exchange or correlation potentials can diverge, for example in the case of the hydrogen 1$s$ type orbital. Figure \ref{fig:1s} shows the SCAN exchange-correlation potential corresponding to the density and kinetic energy density of \replaced{$\psi(r)=\sfrac{e^{-r}}{\sqrt{\pi}}$, the H}{a} $1s$ \deleted{type} orbital, exhibiting the unphysically divergent behavior.

Furness and Sun suggested an alternative iso-orbital indicator function $\beta=\frac{\tau-\tau_W}{\tau+\tau_U}$, that still displays some divergence, but at a significantly smaller rate, therefore in total resulting in a physically well-behaved potential.\cite{Furness:2019ic} In this paper, however, we intend to propose the \emph{least} amount of modification in the SCAN functional form, hence we resorted to regularizing the original iso-orbital indication function.

The worst divergence occurs in the low-density, single-orbital region, where $\alpha \approx 0$, or in case of the 1s orbital example, $\alpha = 0$ exactly. It is partially due to the rapidly decreasing $\tau_U$ in the denominator of $\alpha$, which leads to numerical instabilities at low-density regions in $\alpha$. We propose our first regularization in the kinetic energy density of the uniform electron gas, as $\tau_U' = \tau_U + \tau_{r}$, where $\tau_{r} = 1 \times 10^{-4}$ is a small constant, which only affects $\alpha$ at very low densities.

The second proposed regularization is described as
$\alpha' = \frac{\alpha^3}{\alpha^2 +\alpha_r}$, where $\alpha_r = 1 \times 10^{-3}$ is a small constant, and the regularized iso-orbital indicator function $\alpha'$ only differs from the original $\alpha$ function at small values. However, in the single-orbital region this construction allows vanishing derivatives of $\alpha'$ with respect to $n$, $\nabla n$ and $\tau$, therefore minimizing the interference of the switching construction with the physically motivated parts of the exchange and correlation functional expressions. \added{It should be noted, however, that upon introducing $\tau'$ and $\alpha'$, the exchange energy no longer scales exactly under uniform scaling of the density, although in a practical calculation this effect is expected to remain negligible.}

\begin{figure}[htb]
    \includegraphics[width=\linewidth]{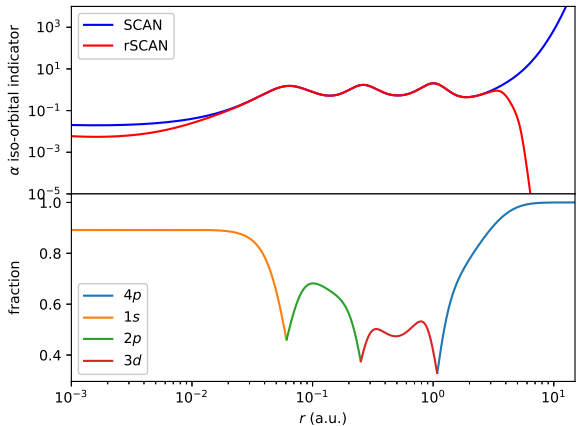}
    \caption{Top panel: iso-orbital indicator function of SCAN and rSCAN, as evaluated on the Kr self-consistent densities computed with the PBE exchange-correlation functional, shown as a function of distance from the nucleus. Bottom panel: At each distance, fraction of the contribution from the highest contributing single orbital to the total electron density for the isolated Kr atom.}
    \label{fig:alpha}
\end{figure}
In low density regions, rSCAN corrects the divergence of derivatives, as well as adjusting the physical interpretation that the iso-orbital indicator provides.
For example, in the case of isolated noble gas atoms, with the exception of helium, the tail of the valence $p$ orbitals tend to dominate far from the nucleus. 
According to the original definition, this results in $\alpha \gg 1$ at greater distances from the nucleus, corresponding to weak bonds\cite{Sun:2015ef}, whereas the regularized $\alpha$ indicator returns to zero.
This is more similar to helium, where $\alpha=0$ everywhere, by construction.
Figure \ref{fig:alpha} compares the original and the regularized iso-orbital indicator functions for the isolated Kr atom, also indicating the proportion of the highest contributing orbital type. 

\subsection{The switching function}
\begin{figure}[htb]
    \includegraphics[width=\linewidth]{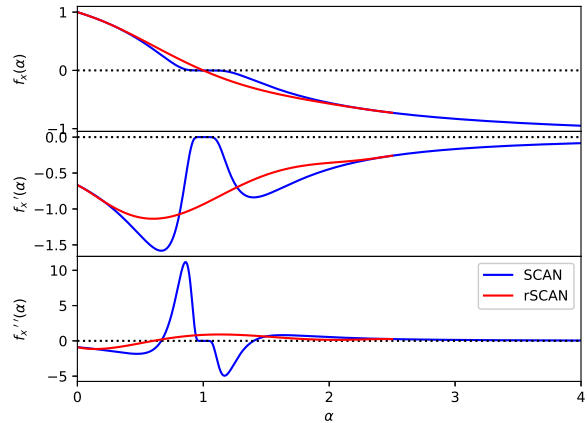}
    \caption{Switching functions and their derivatives used in the exchange functional of SCAN and rSCAN.}
    \label{fig:switch}
\end{figure}

\begin{figure}[htb]
    \includegraphics[width=\linewidth]{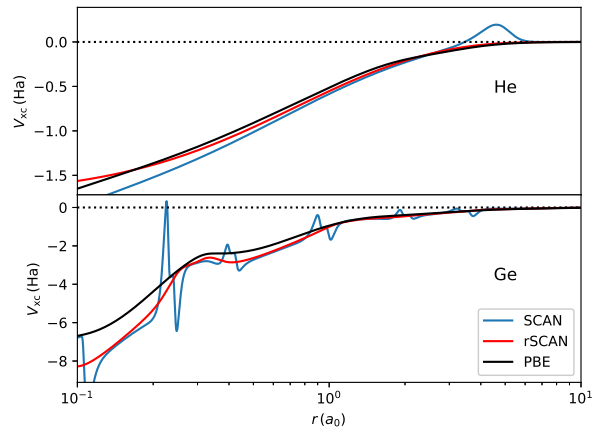}
    \caption{\added{Multiplicative part of the} SCAN and rSCAN exchange-correlation potentials computed using the PBE \added{self-consistent} electronic and kinetic energy densities of isolated He (top) and Ge (bottom) atoms. The PBE result is also shown for reference.}
    \label{fig:atomic}
\end{figure}

The original SCAN functional form includes a switching function, based on the iso-orbital indicator.
The switching function facilitates a smooth transition between limiting cases, which are constructed observing the constraints based on exact \replaced{density functional}{model systems}.
The functional form of the switching function had been carefully selected, and its parameters were fitted such that the resulting exchange and correlation energies reproduce those of accurate model systems.
Even so, the actual form is arbitrary, and we identified the region corresponding to $\alpha \approx 1$ as another source of numerical instability.
Figure \ref{fig:switch} shows the switching function and its first and second derivatives, both contributing to the resulting exchange and correlation potentials. 
The region around $\alpha \approx 1$ is constructed so flat that $f^{(n)}(1)=0$ for every $n$, \replaced{in order to preserve the gradient expansion for the exchange energy in the slowly varying limit. However, this in turn}{which, in turn} introduces severe oscillations in the derivatives at the surrounding region.
These oscillations also manifest in the exchange-correlation potential, as shown \added{for GGA part of the potential} in Figure \ref{fig:atomic} and mentioned in Ref. \onlinecite{Yang:2016ef}.

Our intent is to make minimal changes to the switching function, and we found that replacing the region $0<\alpha<2.5$ by an \replaced{7}{8}-th degree polynomial removes the oscillatory behaviour, while keeping the performance of SCAN similar to the original functional form, \added{although recognizing that we lose the gradient expansion in the slowly varying limit}. We fitted the coefficients\cite{SI} of the polynomials such that the derivatives $f^{(0,1,2)}(0)$, $f^{(0,1,2,3)}(2.5)$ are retained and the additional constraint $f(1)=0$ is satisfied. Figure \ref{fig:switch} compares the original and modified switching functions and their derivatives, demonstrating the improved smoothness, as also evidenced in the practical case of two isolated atoms in Figure \ref{fig:atomic}.

\section{Results}

We implemented rSCAN in the \CASTEP{}\cite{Clark:2005vp} planewave-pseudopotential DFT program and the PySCF quantum chemistry package\cite{Sun:2017kx}. Self-consistent calculations were performed by solving the generalized Kohn-Sham equations iteratively\cite{Yang:2016ef,Perdew:2017cg}. In \CASTEP{}, ultrasoft pseudopotentials were generated on-the-fly, using the methodology we described elsewhere\cite{bartok2019ultrasoft}. We have also pseudized the $\tau$-dependent part, $V_\tau$ of the exchange-correlation potential. In our solid-state calculations, we used Monkhorst-Pack $k$-point grids\cite{Monkhorst:1976ta} with a 0.02~\AA$^{-1}$ (0.014~\AA$^{-1}$ in case of metals) spacing to sample the Brillouin zone, and the \verb+basis_precision : extreme+ setting in \CASTEP{} for the energy cutoff of the planewave basis.
PySCF was used to compute the Ar dimer dissociation energies, using the aug-cc-PVQZ basis set\cite{Woon:1993in} at the standard grid settings. We used \CASTEP{} to optimize the geometry of the water monomer and hexamer configurations, using a cubic box with 15~{\AA} sides, 750~eV planewave cutoff and the $\Gamma$ point in the Brillouin zone.

The parameters in the exchange and correlation switching function of the original SCAN were fitted to reproduce the exchange and correlation energies of isolated Ne, Ar, Kr and Xe atoms, the interaction energies of compressed Ar dimers and the jellium surface exchange-correlation energy. We compared the accuracy of these quantities, with the exception of the jellium surface exchange-correlation energy, and summarized the results in Table \ref{table:XC}. For the relative binding energy curve of the Ar dimer at 1.6~{\AA}, 1.8~{\AA} and 2.0~{\AA}, the mean absolute error of rSCAN is 1.1~kcal/mol, while the figure for the original SCAN was below 1~kcal/mol.

\begin{table}[htb]
\begin{tabular}{ll|cccc}
\toprule
        &       & Ne &  Ar & Kr & Xe \\
\hline
\hline
\midrule
\multirow{3}{*}{$E_\textrm{x}$} & SCAN &  -12.164 & -30.263 & -94.068 & -179.325 \\
                                & rSCAN & -12.163 & -30.298 & -94.199 & -179.632 \\
                                & ref.  & -12.108 & -30.188 & -93.890 & -179.200 \\
\hline
\multirow{3}{*}{$E_\textrm{c}$} & SCAN &  -0.345 & -0.691 & -1.756 & -2.899 \\
                                & rSCAN & -0.345 & -0.695 & -1.768 & -2.914 \\
                                & ref.  & -0.391 & -0.723 & -1.850 & -3.000 \\
\hline
\multirow{3}{*}{$E_\textrm{xc}$} & SCAN &  -12.508 & -30.954 & -95.826 & -182.218 \\
                                 & rSCAN & -12.508 & -30.993 & -95.966 & -182.546 \\
                                 & ref.  & -12.499 & -30.911 & -95.740 & -182.200 \\
\bottomrule
\end{tabular}
\caption{Exchange and correlation energies of isolated noble gas atoms, in hartrees. Original SCAN values are obtained from Ref. \onlinecite{Sun:2015ef}, reference values from Refs. \onlinecite{Becke:1988ka,Chakravorty:1993ux,McCarthy:2011cd}.}
\label{table:XC}
\end{table}

We also benchmarked the rSCAN on some model systems in the literature where results with the original SCAN are available. The set is far from complete, and we note that the literature figures are not consistent: they were obtained by a broad range of codes using different basis sets, in some cases with inconsistent PAW pseudopotentials. However, our results demonstrate that rSCAN has a performance comparable to the original SCAN functional.

Table \ref{table:lattice} lists the lattice constants of a set of simple solids as calculated with rSCAN, and compares them to experiment as well as the original SCAN figures reproduced from the Supplementary Material of Ref.~\onlinecite{Sun:2015ef}, showing good agreement.
A recently published shortcoming of SCAN is the overestimation of magnetic energies of ferromagnetic systems.\cite{Fu:2018cy}
We have found that rSCAN performs similarly, obtaining $m=2.62\,\mu_B$ of the spin moments for bcc iron at the optimized lattice constant of 2.84~{\AA}, in good agreement of the SCAN values presented in Ref. \onlinecite{Fu:2018cy} $m=2.60\,\mu_B$ at the optimized 2.85~{\AA} lattice constant.
\begin{table}[htb]
\begin{tabular}{c|cccccccc}
\toprule
      & Li & Na & Ag & C & Si & SiC & LiF & MgO \\
\hline
Expt. & 3.451 & 4.207 & 4.063 & 3.555 & 5.422 & 
4.348 & 3.974 & 4.188\\
SCAN  & 3.460 & 4.190 & 4.079 & 3.550 & 5.424 &
4.349 & 3.980 & 4.206\\
rSCAN & 3.453 & 4.197 & 4.039 & 3.555 & 5.441 &
4.353 & 3.964 & 4.200
\end{tabular}
\caption{Equilibrium lattice constants (\AA) of a selection of metallic and semiconductor solids (a subset of ``LC20'' in Ref.\onlinecite{Sun:2015ef}), computed using the rSCAN functional. Experimental values, corrected for zero point anharmonic expansion, were taken from Ref. \onlinecite{Hao:2012da}, and reference SCAN values from Ref. \onlinecite{Sun:2015ef}.}
\label{table:lattice}
\end{table}


Interaction energies of water systems are a very strict test of density functionals, and the original SCAN functional performs remarkably well, predicting the correct energetic ordering of ice polymorphs and water hexamer conformations. With the rSCAN, the water monomer geometry is very close to that of the original SCAN and the dipole moments of the isolated molecule are also in close agreement. We have also calculated the dissociation energies of four low-energy water hexamers, as shown in Table~\ref{tab:water}, recovering the same energetic ordering as predicted by CCSD(T)\cite{Santra:2008jw} and SCAN, and somewhat improving the absolute values of the energies.
\begin{table}[]
    \begin{tabular}{c|cccc|ccc}
    & prism & cage & book & chair & 
    $r_\textrm{OH}$ & $\theta_\textrm{HOH}$ & $\mu$ \\
\hline\hline
 ref. & 348 & 346 & 339 & 332 & 0.957 & 104.5\degree & 1.855 \\
 SCAN & 377 & 376 & 370 & 360 & 0.961 & 104.5\degree &
 1.847 \\
 rSCAN & 359 & 358 & 356 & 348 & 0.959 & 104.4\degree &
 1.847
    \end{tabular}
    \caption{Dissociation energies (meV/monomer) of a few low-energy water hexamers conformations, the equilibrium bond length ({\AA}), bond angle and dipole moment (Debye) of the water molecule. Reference hexamer dissociation values are computed by CCSD(T)\cite{Santra:2008jw}, while the geometry of the water molecule is from Ref.~\onlinecite{Benedict:1956id} and its dipole moment from Ref.~\onlinecite{Dyke:1973bv}. SCAN values were obtained from Ref.~\onlinecite{Sun:2016jp}.}
    \label{tab:water}
\end{table}

\section{Conclusions}
Exchange-correlation functionals based on the meta-Generalized Gradient Approximation have become increasingly successful, but their implementation in solid-state DFT packages lags behind the theoretical developments. We have implemented the SCAN mGGA functional in a plane-wave DFT program, using ultrasoft pseudopotentials generated with the same functional, and solving the electronic problem self-consistently via the generalized Kohn-Sham scheme. To achieve this it was necessary to introduce a regularized form of the SCAN functional that has an improved numerical stability while retaining the accuracy of the original form. We note that the few adjustable parameters which we imported from SCAN may be re-optimized to further improve the performance, but that is outside of the scope of our current work.
Our benchmark calculations illustrate that the proposed rSCAN functional remains transferable and accurate for a broad range of solid state and molecular systems. rSCAN will make the generation of pseudopotential and PAW datasets more straightforward in other packages, while its improved smoothness properties should improve the stability of any DFT implementation where the exchange-correlation functionals need to be represented on a grid.

\section*{Supplementary Material}
\added{See Supplementary Material\cite{SI} for the numerical values of the polynomial coefficients of the modified exchange and correlation switching functions.}

\begin{acknowledgments}
We would like to thank Chris Pickard, Philip Hasnip and Dominik Jochym for useful discussions. Both authors acknowledge support from the Collaborative Computational Project for NMR Crystallography (CCP-NC) and UKCP Consortium, both funded by the Engineering and Physical Sciences Research Council (EPSRC) under grant numbers EP/M022501/1 and EP/P022561/1, respectively. Computing resources were provided by the STFC Scientific Computing Department's SCARF cluster.
\end{acknowledgments}

\bibliography{rSCAN}

\begin{thebibliography}{34}%
\makeatletter
\providecommand \@ifxundefined [1]{%
 \@ifx{#1\undefined}
}%
\providecommand \@ifnum [1]{%
 \ifnum #1\expandafter \@firstoftwo
 \else \expandafter \@secondoftwo
 \fi
}%
\providecommand \@ifx [1]{%
 \ifx #1\expandafter \@firstoftwo
 \else \expandafter \@secondoftwo
 \fi
}%
\providecommand \natexlab [1]{#1}%
\providecommand \enquote  [1]{``#1''}%
\providecommand \bibnamefont  [1]{#1}%
\providecommand \bibfnamefont [1]{#1}%
\providecommand \citenamefont [1]{#1}%
\providecommand \href@noop [0]{\@secondoftwo}%
\providecommand \href [0]{\begingroup \@sanitize@url \@href}%
\providecommand \@href[1]{\@@startlink{#1}\@@href}%
\providecommand \@@href[1]{\endgroup#1\@@endlink}%
\providecommand \@sanitize@url [0]{\catcode `\\12\catcode `\$12\catcode
  `\&12\catcode `\#12\catcode `\^12\catcode `\_12\catcode `\%12\relax}%
\providecommand \@@startlink[1]{}%
\providecommand \@@endlink[0]{}%
\providecommand \url  [0]{\begingroup\@sanitize@url \@url }%
\providecommand \@url [1]{\endgroup\@href {#1}{\urlprefix }}%
\providecommand \urlprefix  [0]{URL }%
\providecommand \Eprint [0]{\href }%
\providecommand \doibase [0]{http://dx.doi.org/}%
\providecommand \selectlanguage [0]{\@gobble}%
\providecommand \bibinfo  [0]{\@secondoftwo}%
\providecommand \bibfield  [0]{\@secondoftwo}%
\providecommand \translation [1]{[#1]}%
\providecommand \BibitemOpen [0]{}%
\providecommand \bibitemStop [0]{}%
\providecommand \bibitemNoStop [0]{.\EOS\space}%
\providecommand \EOS [0]{\spacefactor3000\relax}%
\providecommand \BibitemShut  [1]{\csname bibitem#1\endcsname}%
\let\auto@bib@innerbib\@empty
\bibitem [{\citenamefont {Jones}(2015)}]{Jones:2015gg}%
  \BibitemOpen
  \bibfield  {author} {\bibinfo {author} {\bibfnamefont {R.~O.}\ \bibnamefont
  {Jones}},\ }\href@noop {} {\bibfield  {journal} {\bibinfo  {journal} {Rev.
  Mod. Phys.}\ }\textbf {\bibinfo {volume} {87}},\ \bibinfo {pages} {897}
  (\bibinfo {year} {2015})}\BibitemShut {NoStop}%
\bibitem [{\citenamefont {Kohn}\ and\ \citenamefont
  {Sham}(1965)}]{Kohn:1965uia}%
  \BibitemOpen
  \bibfield  {author} {\bibinfo {author} {\bibfnamefont {W.}~\bibnamefont
  {Kohn}}\ and\ \bibinfo {author} {\bibfnamefont {L.}~\bibnamefont {Sham}},\
  }\href@noop {} {\bibfield  {journal} {\bibinfo  {journal} {Phys. Rev.}\
  }\textbf {\bibinfo {volume} {140}},\ \bibinfo {pages} {A1133} (\bibinfo
  {year} {1965})}\BibitemShut {NoStop}%
\bibitem [{\citenamefont {Perdew}\ \emph {et~al.}(1996)\citenamefont {Perdew},
  \citenamefont {Burke},\ and\ \citenamefont {Ernzerhof}}]{Perdew:1996gs}%
  \BibitemOpen
  \bibfield  {author} {\bibinfo {author} {\bibfnamefont {J.~P.}\ \bibnamefont
  {Perdew}}, \bibinfo {author} {\bibfnamefont {K.}~\bibnamefont {Burke}}, \
  and\ \bibinfo {author} {\bibfnamefont {M.}~\bibnamefont {Ernzerhof}},\
  }\href@noop {} {\bibfield  {journal} {\bibinfo  {journal} {Phys. Rev. Lett.}\
  }\textbf {\bibinfo {volume} {77}},\ \bibinfo {pages} {3865} (\bibinfo {year}
  {1996})}\BibitemShut {NoStop}%
\bibitem [{\citenamefont {Sun}\ \emph {et~al.}(2016)\citenamefont {Sun},
  \citenamefont {Remsing}, \citenamefont {Zhang}, \citenamefont {Sun},
  \citenamefont {Ruzsinszky}, \citenamefont {Peng}, \citenamefont {Yang},
  \citenamefont {Paul}, \citenamefont {Waghmare}, \citenamefont {Wu},
  \citenamefont {Klein},\ and\ \citenamefont {Perdew}}]{Sun:2016jp}%
  \BibitemOpen
  \bibfield  {author} {\bibinfo {author} {\bibfnamefont {J.}~\bibnamefont
  {Sun}}, \bibinfo {author} {\bibfnamefont {R.~C.}\ \bibnamefont {Remsing}},
  \bibinfo {author} {\bibfnamefont {Y.}~\bibnamefont {Zhang}}, \bibinfo
  {author} {\bibfnamefont {Z.}~\bibnamefont {Sun}}, \bibinfo {author}
  {\bibfnamefont {A.}~\bibnamefont {Ruzsinszky}}, \bibinfo {author}
  {\bibfnamefont {H.}~\bibnamefont {Peng}}, \bibinfo {author} {\bibfnamefont
  {Z.}~\bibnamefont {Yang}}, \bibinfo {author} {\bibfnamefont {A.}~\bibnamefont
  {Paul}}, \bibinfo {author} {\bibfnamefont {U.}~\bibnamefont {Waghmare}},
  \bibinfo {author} {\bibfnamefont {X.}~\bibnamefont {Wu}}, \bibinfo {author}
  {\bibfnamefont {M.~L.}\ \bibnamefont {Klein}}, \ and\ \bibinfo {author}
  {\bibfnamefont {J.~P.}\ \bibnamefont {Perdew}},\ }\href@noop {} {\bibfield
  {journal} {\bibinfo  {journal} {Nat. Chem.}\ }\textbf {\bibinfo {volume}
  {8}},\ \bibinfo {pages} {831} (\bibinfo {year} {2016})}\BibitemShut {NoStop}%
\bibitem [{\citenamefont {Chen}\ \emph {et~al.}(2017)\citenamefont {Chen},
  \citenamefont {Ko}, \citenamefont {Remsing}, \citenamefont
  {Calegari~Andrade}, \citenamefont {Santra}, \citenamefont {Sun},
  \citenamefont {Selloni}, \citenamefont {Car}, \citenamefont {Klein},
  \citenamefont {Perdew},\ and\ \citenamefont {Wu}}]{Chen:2017jn}%
  \BibitemOpen
  \bibfield  {author} {\bibinfo {author} {\bibfnamefont {M.}~\bibnamefont
  {Chen}}, \bibinfo {author} {\bibfnamefont {H.-Y.}\ \bibnamefont {Ko}},
  \bibinfo {author} {\bibfnamefont {R.~C.}\ \bibnamefont {Remsing}}, \bibinfo
  {author} {\bibfnamefont {M.~F.}\ \bibnamefont {Calegari~Andrade}}, \bibinfo
  {author} {\bibfnamefont {B.}~\bibnamefont {Santra}}, \bibinfo {author}
  {\bibfnamefont {Z.}~\bibnamefont {Sun}}, \bibinfo {author} {\bibfnamefont
  {A.}~\bibnamefont {Selloni}}, \bibinfo {author} {\bibfnamefont
  {R.}~\bibnamefont {Car}}, \bibinfo {author} {\bibfnamefont {M.~L.}\
  \bibnamefont {Klein}}, \bibinfo {author} {\bibfnamefont {J.~P.}\ \bibnamefont
  {Perdew}}, \ and\ \bibinfo {author} {\bibfnamefont {X.}~\bibnamefont {Wu}},\
  }\href@noop {} {\bibfield  {journal} {\bibinfo  {journal} {Proc. Natl. Acad.
  Sci.}\ }\textbf {\bibinfo {volume} {114}},\ \bibinfo {pages} {10846}
  (\bibinfo {year} {2017})}\BibitemShut {NoStop}%
\bibitem [{\citenamefont {Remsing}\ \emph {et~al.}(2017)\citenamefont
  {Remsing}, \citenamefont {Klein},\ and\ \citenamefont
  {Sun}}]{Remsing:2017fy}%
  \BibitemOpen
  \bibfield  {author} {\bibinfo {author} {\bibfnamefont {R.~C.}\ \bibnamefont
  {Remsing}}, \bibinfo {author} {\bibfnamefont {M.~L.}\ \bibnamefont {Klein}},
  \ and\ \bibinfo {author} {\bibfnamefont {J.}~\bibnamefont {Sun}},\
  }\href@noop {} {\bibfield  {journal} {\bibinfo  {journal} {Phys. Rev. B}\
  }\textbf {\bibinfo {volume} {96}},\ \bibinfo {pages} {831} (\bibinfo {year}
  {2017})}\BibitemShut {NoStop}%
\bibitem [{\citenamefont {Gautam}\ and\ \citenamefont
  {Carter}(2018)}]{Gautam:2018gx}%
  \BibitemOpen
  \bibfield  {author} {\bibinfo {author} {\bibfnamefont {G.~S.}\ \bibnamefont
  {Gautam}}\ and\ \bibinfo {author} {\bibfnamefont {E.~A.}\ \bibnamefont
  {Carter}},\ }\href@noop {} {\bibfield  {journal} {\bibinfo  {journal} {Phys.
  Rev. Mat.}\ }\textbf {\bibinfo {volume} {2}},\ \bibinfo {pages} {095401}
  (\bibinfo {year} {2018})}\BibitemShut {NoStop}%
\bibitem [{elk()}]{elk}%
  \BibitemOpen
  \href@noop {} {}\bibinfo {note}
  {\texttt{http://elk.sourceforge.net/}}\BibitemShut {NoStop}%
\bibitem [{\citenamefont {Blaha}\ \emph {et~al.}(2018)\citenamefont {Blaha},
  \citenamefont {Schwarz}, \citenamefont {Madsen}, \citenamefont {Kvasnicka},
  \citenamefont {Luitz}, \citenamefont {R.}, \citenamefont {F.},\ and\
  \citenamefont {D.}}]{wien2k}%
  \BibitemOpen
  \bibfield  {author} {\bibinfo {author} {\bibfnamefont {P.}~\bibnamefont
  {Blaha}}, \bibinfo {author} {\bibfnamefont {K.}~\bibnamefont {Schwarz}},
  \bibinfo {author} {\bibfnamefont {G.~K.~H.}\ \bibnamefont {Madsen}}, \bibinfo
  {author} {\bibfnamefont {D.}~\bibnamefont {Kvasnicka}}, \bibinfo {author}
  {\bibfnamefont {J.}~\bibnamefont {Luitz}}, \bibinfo {author} {\bibfnamefont
  {L.}~\bibnamefont {R.}}, \bibinfo {author} {\bibfnamefont {T.}~\bibnamefont
  {F.}}, \ and\ \bibinfo {author} {\bibfnamefont {M.~L.}\ \bibnamefont {D.}},\
  }\href@noop {} {\emph {\bibinfo {title} {{WIEN2K}, {A}n {A}ugmented {P}lane
  {W}ave + {L}ocal {O}rbitals {P}rogram for {C}alculating {C}rystal
  {P}roperties}}}\ (\bibinfo  {publisher} {{K}arlheinz Schwarz, Techn.
  Universit\"{a}t Wien, Austria},\ \bibinfo {year} {2018})\BibitemShut
  {NoStop}%
\bibitem [{\citenamefont {Yao}\ and\ \citenamefont {Kanai}(2017)}]{Yao:2017gn}%
  \BibitemOpen
  \bibfield  {author} {\bibinfo {author} {\bibfnamefont {Y.}~\bibnamefont
  {Yao}}\ and\ \bibinfo {author} {\bibfnamefont {Y.}~\bibnamefont {Kanai}},\
  }\href@noop {} {\bibfield  {journal} {\bibinfo  {journal} {J. Chem. Phys.}\
  }\textbf {\bibinfo {volume} {146}},\ \bibinfo {pages} {224105} (\bibinfo
  {year} {2017})}\BibitemShut {NoStop}%
\bibitem [{\citenamefont {Fu}\ and\ \citenamefont {Singh}(2018)}]{Fu:2018cy}%
  \BibitemOpen
  \bibfield  {author} {\bibinfo {author} {\bibfnamefont {Y.}~\bibnamefont
  {Fu}}\ and\ \bibinfo {author} {\bibfnamefont {D.~J.}\ \bibnamefont {Singh}},\
  }\href@noop {} {\bibfield  {journal} {\bibinfo  {journal} {Phys. Rev. Lett.}\
  }\textbf {\bibinfo {volume} {121}},\ \bibinfo {pages} {207201} (\bibinfo
  {year} {2018})}\BibitemShut {NoStop}%
\bibitem [{\citenamefont {Janthon}\ \emph {et~al.}(2014)\citenamefont
  {Janthon}, \citenamefont {Luo}, \citenamefont {Kozlov}, \citenamefont
  {Vi{\~n}es}, \citenamefont {Limtrakul}, \citenamefont {Truhlar},\ and\
  \citenamefont {Illas}}]{Janthon:2014dr}%
  \BibitemOpen
  \bibfield  {author} {\bibinfo {author} {\bibfnamefont {P.}~\bibnamefont
  {Janthon}}, \bibinfo {author} {\bibfnamefont {S.~A.}\ \bibnamefont {Luo}},
  \bibinfo {author} {\bibfnamefont {S.~M.}\ \bibnamefont {Kozlov}}, \bibinfo
  {author} {\bibfnamefont {F.}~\bibnamefont {Vi{\~n}es}}, \bibinfo {author}
  {\bibfnamefont {J.}~\bibnamefont {Limtrakul}}, \bibinfo {author}
  {\bibfnamefont {D.~G.}\ \bibnamefont {Truhlar}}, \ and\ \bibinfo {author}
  {\bibfnamefont {F.}~\bibnamefont {Illas}},\ }\href@noop {} {\bibfield
  {journal} {\bibinfo  {journal} {J. Chem. Theory Comp.}\ }\textbf {\bibinfo
  {volume} {10}},\ \bibinfo {pages} {3832} (\bibinfo {year}
  {2014})}\BibitemShut {NoStop}%
\bibitem [{\citenamefont {Dorner}\ \emph {et~al.}(2018)\citenamefont {Dorner},
  \citenamefont {Sukurma}, \citenamefont {Dellago},\ and\ \citenamefont
  {Kresse}}]{Dorner:2018cp}%
  \BibitemOpen
  \bibfield  {author} {\bibinfo {author} {\bibfnamefont {F.}~\bibnamefont
  {Dorner}}, \bibinfo {author} {\bibfnamefont {Z.}~\bibnamefont {Sukurma}},
  \bibinfo {author} {\bibfnamefont {C.}~\bibnamefont {Dellago}}, \ and\
  \bibinfo {author} {\bibfnamefont {G.}~\bibnamefont {Kresse}},\ }\href
  {\doibase 10.1103/PhysRevLett.121.195701} {\bibfield  {journal} {\bibinfo
  {journal} {Phys. Rev. Lett.}\ }\textbf {\bibinfo {volume} {121}},\ \bibinfo
  {pages} {195701} (\bibinfo {year} {2018})}\BibitemShut {NoStop}%
\bibitem [{\citenamefont {Bonati}\ and\ \citenamefont
  {Parrinello}(2018)}]{Bonati:2018iu}%
  \BibitemOpen
  \bibfield  {author} {\bibinfo {author} {\bibfnamefont {L.}~\bibnamefont
  {Bonati}}\ and\ \bibinfo {author} {\bibfnamefont {M.}~\bibnamefont
  {Parrinello}},\ }\href@noop {} {\bibfield  {journal} {\bibinfo  {journal}
  {Phys. Rev. Lett.}\ }\textbf {\bibinfo {volume} {121}},\ \bibinfo {pages}
  {265701} (\bibinfo {year} {2018})}\BibitemShut {NoStop}%
\bibitem [{\citenamefont {Isaacs}\ and\ \citenamefont
  {Wolverton}(2018)}]{Isaacs:2018hm}%
  \BibitemOpen
  \bibfield  {author} {\bibinfo {author} {\bibfnamefont {E.~B.}\ \bibnamefont
  {Isaacs}}\ and\ \bibinfo {author} {\bibfnamefont {C.}~\bibnamefont
  {Wolverton}},\ }\href@noop {} {\bibfield  {journal} {\bibinfo  {journal}
  {Phys. Rev. Mat.}\ }\textbf {\bibinfo {volume} {2}},\ \bibinfo {pages}
  {063801} (\bibinfo {year} {2018})}\BibitemShut {NoStop}%
\bibitem [{\citenamefont {Fuchs}\ \emph {et~al.}(1998)\citenamefont {Fuchs},
  \citenamefont {Bockstedte}, \citenamefont {Pehlke},\ and\ \citenamefont
  {Scheffler}}]{Fuchs:1998cf}%
  \BibitemOpen
  \bibfield  {author} {\bibinfo {author} {\bibfnamefont {M.}~\bibnamefont
  {Fuchs}}, \bibinfo {author} {\bibfnamefont {M.}~\bibnamefont {Bockstedte}},
  \bibinfo {author} {\bibfnamefont {E.}~\bibnamefont {Pehlke}}, \ and\ \bibinfo
  {author} {\bibfnamefont {M.}~\bibnamefont {Scheffler}},\ }\href@noop {}
  {\bibfield  {journal} {\bibinfo  {journal} {Phys. Rev. B}\ }\textbf {\bibinfo
  {volume} {57}},\ \bibinfo {pages} {2134} (\bibinfo {year}
  {1998})}\BibitemShut {NoStop}%
\bibitem [{\citenamefont {Bart{\'o}k}\ and\ \citenamefont
  {Yates}(2019)}]{bartok2019ultrasoft}%
  \BibitemOpen
  \bibfield  {author} {\bibinfo {author} {\bibfnamefont {A.~P.}\ \bibnamefont
  {Bart{\'o}k}}\ and\ \bibinfo {author} {\bibfnamefont {J.~R.}\ \bibnamefont
  {Yates}},\ }\href@noop {} {\bibfield  {journal} {\bibinfo  {journal} {arXiv}\
  ,\ \bibinfo {pages} {1901.11301}} (\bibinfo {year} {2019})}\BibitemShut
  {NoStop}%
\bibitem [{\citenamefont {Yang}\ \emph {et~al.}(2016)\citenamefont {Yang},
  \citenamefont {Peng}, \citenamefont {Sun},\ and\ \citenamefont
  {Perdew}}]{Yang:2016ef}%
  \BibitemOpen
  \bibfield  {author} {\bibinfo {author} {\bibfnamefont {Z.-h.}\ \bibnamefont
  {Yang}}, \bibinfo {author} {\bibfnamefont {H.}~\bibnamefont {Peng}}, \bibinfo
  {author} {\bibfnamefont {J.}~\bibnamefont {Sun}}, \ and\ \bibinfo {author}
  {\bibfnamefont {J.~P.}\ \bibnamefont {Perdew}},\ }\href@noop {} {\bibfield
  {journal} {\bibinfo  {journal} {Phys. Rev. B}\ }\textbf {\bibinfo {volume}
  {93}},\ \bibinfo {pages} {205205} (\bibinfo {year} {2016})}\BibitemShut
  {NoStop}%
\bibitem [{\citenamefont {Furness}\ and\ \citenamefont
  {Sun}(2019)}]{Furness:2019ic}%
  \BibitemOpen
  \bibfield  {author} {\bibinfo {author} {\bibfnamefont {J.~W.}\ \bibnamefont
  {Furness}}\ and\ \bibinfo {author} {\bibfnamefont {J.}~\bibnamefont {Sun}},\
  }\href@noop {} {\bibfield  {journal} {\bibinfo  {journal} {Phys. Rev. B}\
  }\textbf {\bibinfo {volume} {99}},\ \bibinfo {pages} {041119} (\bibinfo
  {year} {2019})}\BibitemShut {NoStop}%
\bibitem [{\citenamefont {Sun}\ \emph {et~al.}(2015)\citenamefont {Sun},
  \citenamefont {Ruzsinszky},\ and\ \citenamefont {Perdew}}]{Sun:2015ef}%
  \BibitemOpen
  \bibfield  {author} {\bibinfo {author} {\bibfnamefont {J.}~\bibnamefont
  {Sun}}, \bibinfo {author} {\bibfnamefont {A.}~\bibnamefont {Ruzsinszky}}, \
  and\ \bibinfo {author} {\bibfnamefont {J.~P.}\ \bibnamefont {Perdew}},\
  }\href@noop {} {\bibfield  {journal} {\bibinfo  {journal} {Phys. Rev. Lett.}\
  }\textbf {\bibinfo {volume} {115}},\ \bibinfo {pages} {036402} (\bibinfo
  {year} {2015})}\BibitemShut {NoStop}%
\bibitem [{\citenamefont {Sun}\ \emph {et~al.}(2013)\citenamefont {Sun},
  \citenamefont {Haunschild}, \citenamefont {Xiao}, \citenamefont {Bulik},
  \citenamefont {Scuseria},\ and\ \citenamefont {Perdew}}]{Sun:2013ku}%
  \BibitemOpen
  \bibfield  {author} {\bibinfo {author} {\bibfnamefont {J.}~\bibnamefont
  {Sun}}, \bibinfo {author} {\bibfnamefont {R.}~\bibnamefont {Haunschild}},
  \bibinfo {author} {\bibfnamefont {B.}~\bibnamefont {Xiao}}, \bibinfo {author}
  {\bibfnamefont {I.~W.}\ \bibnamefont {Bulik}}, \bibinfo {author}
  {\bibfnamefont {G.~E.}\ \bibnamefont {Scuseria}}, \ and\ \bibinfo {author}
  {\bibfnamefont {J.~P.}\ \bibnamefont {Perdew}},\ }\href@noop {} {\bibfield
  {journal} {\bibinfo  {journal} {J. Chem. Phys.}\ }\textbf {\bibinfo {volume}
  {138}},\ \bibinfo {pages} {044113} (\bibinfo {year} {2013})}\BibitemShut
  {NoStop}%
\bibitem [{SI()}]{SI}%
  \BibitemOpen
  \href@noop {} {}\bibinfo {note} {See Supplementary Information.}\BibitemShut
  {Stop}%
\bibitem [{\citenamefont {Clark}\ \emph {et~al.}(2005)\citenamefont {Clark},
  \citenamefont {Segall}, \citenamefont {Pickard}, \citenamefont {Hasnip},
  \citenamefont {Probert}, \citenamefont {Refson},\ and\ \citenamefont
  {Payne}}]{Clark:2005vp}%
  \BibitemOpen
  \bibfield  {author} {\bibinfo {author} {\bibfnamefont {S.}~\bibnamefont
  {Clark}}, \bibinfo {author} {\bibfnamefont {M.}~\bibnamefont {Segall}},
  \bibinfo {author} {\bibfnamefont {C.}~\bibnamefont {Pickard}}, \bibinfo
  {author} {\bibfnamefont {P.}~\bibnamefont {Hasnip}}, \bibinfo {author}
  {\bibfnamefont {M.}~\bibnamefont {Probert}}, \bibinfo {author} {\bibfnamefont
  {K.}~\bibnamefont {Refson}}, \ and\ \bibinfo {author} {\bibfnamefont
  {M.}~\bibnamefont {Payne}},\ }\href@noop {} {\bibfield  {journal} {\bibinfo
  {journal} {Z. Kristall.}\ }\textbf {\bibinfo {volume} {220}},\ \bibinfo
  {pages} {567} (\bibinfo {year} {2005})}\BibitemShut {NoStop}%
\bibitem [{\citenamefont {Sun}\ \emph {et~al.}(2017)\citenamefont {Sun},
  \citenamefont {Berkelbach}, \citenamefont {Blunt}, \citenamefont {Booth},
  \citenamefont {Guo}, \citenamefont {Li}, \citenamefont {Liu}, \citenamefont
  {McClain}, \citenamefont {Sayfutyarova}, \citenamefont {Sharma},
  \citenamefont {Wouters},\ and\ \citenamefont {Chan}}]{Sun:2017kx}%
  \BibitemOpen
  \bibfield  {author} {\bibinfo {author} {\bibfnamefont {Q.}~\bibnamefont
  {Sun}}, \bibinfo {author} {\bibfnamefont {T.~C.}\ \bibnamefont {Berkelbach}},
  \bibinfo {author} {\bibfnamefont {N.~S.}\ \bibnamefont {Blunt}}, \bibinfo
  {author} {\bibfnamefont {G.~H.}\ \bibnamefont {Booth}}, \bibinfo {author}
  {\bibfnamefont {S.}~\bibnamefont {Guo}}, \bibinfo {author} {\bibfnamefont
  {Z.}~\bibnamefont {Li}}, \bibinfo {author} {\bibfnamefont {J.}~\bibnamefont
  {Liu}}, \bibinfo {author} {\bibfnamefont {J.~D.}\ \bibnamefont {McClain}},
  \bibinfo {author} {\bibfnamefont {E.~R.}\ \bibnamefont {Sayfutyarova}},
  \bibinfo {author} {\bibfnamefont {S.}~\bibnamefont {Sharma}}, \bibinfo
  {author} {\bibfnamefont {S.}~\bibnamefont {Wouters}}, \ and\ \bibinfo
  {author} {\bibfnamefont {G.~K.-L.}\ \bibnamefont {Chan}},\ }\href@noop {}
  {\bibfield  {journal} {\bibinfo  {journal} {Wiley Interdiscip. Rev. Comput.
  Mol. Sci.}\ }\textbf {\bibinfo {volume} {8}},\ \bibinfo {pages} {e1340}
  (\bibinfo {year} {2017})}\BibitemShut {NoStop}%
\bibitem [{\citenamefont {Perdew}\ \emph {et~al.}(2017)\citenamefont {Perdew},
  \citenamefont {Yang}, \citenamefont {Burke}, \citenamefont {Yang},
  \citenamefont {Gross}, \citenamefont {Scheffler}, \citenamefont {Scuseria},
  \citenamefont {Henderson}, \citenamefont {Zhang}, \citenamefont {Ruzsinszky},
  \citenamefont {Peng}, \citenamefont {Sun}, \citenamefont {Trushin},\ and\
  \citenamefont {G{\"o}rling}}]{Perdew:2017cg}%
  \BibitemOpen
  \bibfield  {author} {\bibinfo {author} {\bibfnamefont {J.~P.}\ \bibnamefont
  {Perdew}}, \bibinfo {author} {\bibfnamefont {W.}~\bibnamefont {Yang}},
  \bibinfo {author} {\bibfnamefont {K.}~\bibnamefont {Burke}}, \bibinfo
  {author} {\bibfnamefont {Z.}~\bibnamefont {Yang}}, \bibinfo {author}
  {\bibfnamefont {E.~K.~U.}\ \bibnamefont {Gross}}, \bibinfo {author}
  {\bibfnamefont {M.}~\bibnamefont {Scheffler}}, \bibinfo {author}
  {\bibfnamefont {G.~E.}\ \bibnamefont {Scuseria}}, \bibinfo {author}
  {\bibfnamefont {T.~M.}\ \bibnamefont {Henderson}}, \bibinfo {author}
  {\bibfnamefont {I.~Y.}\ \bibnamefont {Zhang}}, \bibinfo {author}
  {\bibfnamefont {A.}~\bibnamefont {Ruzsinszky}}, \bibinfo {author}
  {\bibfnamefont {H.}~\bibnamefont {Peng}}, \bibinfo {author} {\bibfnamefont
  {J.}~\bibnamefont {Sun}}, \bibinfo {author} {\bibfnamefont {E.}~\bibnamefont
  {Trushin}}, \ and\ \bibinfo {author} {\bibfnamefont {A.}~\bibnamefont
  {G{\"o}rling}},\ }\href@noop {} {\bibfield  {journal} {\bibinfo  {journal}
  {Proc. Natl. Acad. Sci.}\ }\textbf {\bibinfo {volume} {114}},\ \bibinfo
  {pages} {2801} (\bibinfo {year} {2017})}\BibitemShut {NoStop}%
\bibitem [{\citenamefont {Monkhorst}\ and\ \citenamefont
  {Pack}(1976)}]{Monkhorst:1976ta}%
  \BibitemOpen
  \bibfield  {author} {\bibinfo {author} {\bibfnamefont {H.}~\bibnamefont
  {Monkhorst}}\ and\ \bibinfo {author} {\bibfnamefont {J.}~\bibnamefont
  {Pack}},\ }\href@noop {} {\bibfield  {journal} {\bibinfo  {journal} {Phys.
  Rev. B}\ }\textbf {\bibinfo {volume} {13}},\ \bibinfo {pages} {5188}
  (\bibinfo {year} {1976})}\BibitemShut {NoStop}%
\bibitem [{\citenamefont {Woon}\ and\ \citenamefont
  {Dunning~Jr.}(1993)}]{Woon:1993in}%
  \BibitemOpen
  \bibfield  {author} {\bibinfo {author} {\bibfnamefont {D.~E.}\ \bibnamefont
  {Woon}}\ and\ \bibinfo {author} {\bibfnamefont {T.~H.}\ \bibnamefont
  {Dunning~Jr.}},\ }\href@noop {} {\bibfield  {journal} {\bibinfo  {journal}
  {J. Chem. Phys.}\ }\textbf {\bibinfo {volume} {98}},\ \bibinfo {pages} {1358}
  (\bibinfo {year} {1993})}\BibitemShut {NoStop}%
\bibitem [{\citenamefont {Becke}(1988)}]{Becke:1988ka}%
  \BibitemOpen
  \bibfield  {author} {\bibinfo {author} {\bibfnamefont {A.~D.}\ \bibnamefont
  {Becke}},\ }\href@noop {} {\bibfield  {journal} {\bibinfo  {journal} {Phys.
  Rev. A}\ }\textbf {\bibinfo {volume} {38}},\ \bibinfo {pages} {3098}
  (\bibinfo {year} {1988})}\BibitemShut {NoStop}%
\bibitem [{\citenamefont {Chakravorty}\ \emph {et~al.}(1993)\citenamefont
  {Chakravorty}, \citenamefont {Gwaltney}, \citenamefont {Davidson},
  \citenamefont {Parpia},\ and\ \citenamefont {Fischer}}]{Chakravorty:1993ux}%
  \BibitemOpen
  \bibfield  {author} {\bibinfo {author} {\bibfnamefont {S.~J.}\ \bibnamefont
  {Chakravorty}}, \bibinfo {author} {\bibfnamefont {S.~R.}\ \bibnamefont
  {Gwaltney}}, \bibinfo {author} {\bibfnamefont {E.~R.}\ \bibnamefont
  {Davidson}}, \bibinfo {author} {\bibfnamefont {F.~A.}\ \bibnamefont
  {Parpia}}, \ and\ \bibinfo {author} {\bibfnamefont {C.~F.}\ \bibnamefont
  {Fischer}},\ }\href@noop {} {\bibfield  {journal} {\bibinfo  {journal} {Phys.
  Rev. A}\ }\textbf {\bibinfo {volume} {47}},\ \bibinfo {pages} {3649}
  (\bibinfo {year} {1993})}\BibitemShut {NoStop}%
\bibitem [{\citenamefont {McCarthy}\ and\ \citenamefont
  {Thakkar}(2011)}]{McCarthy:2011cd}%
  \BibitemOpen
  \bibfield  {author} {\bibinfo {author} {\bibfnamefont {S.~P.}\ \bibnamefont
  {McCarthy}}\ and\ \bibinfo {author} {\bibfnamefont {A.~J.}\ \bibnamefont
  {Thakkar}},\ }\href@noop {} {\bibfield  {journal} {\bibinfo  {journal} {J.
  Chem. Phys.}\ }\textbf {\bibinfo {volume} {134}},\ \bibinfo {pages} {044102}
  (\bibinfo {year} {2011})}\BibitemShut {NoStop}%
\bibitem [{\citenamefont {Hao}\ \emph {et~al.}(2012)\citenamefont {Hao},
  \citenamefont {Fang}, \citenamefont {Sun}, \citenamefont {Csonka},
  \citenamefont {Philipsen},\ and\ \citenamefont {Perdew}}]{Hao:2012da}%
  \BibitemOpen
  \bibfield  {author} {\bibinfo {author} {\bibfnamefont {P.}~\bibnamefont
  {Hao}}, \bibinfo {author} {\bibfnamefont {Y.}~\bibnamefont {Fang}}, \bibinfo
  {author} {\bibfnamefont {J.}~\bibnamefont {Sun}}, \bibinfo {author}
  {\bibfnamefont {G.~I.}\ \bibnamefont {Csonka}}, \bibinfo {author}
  {\bibfnamefont {P.~H.~T.}\ \bibnamefont {Philipsen}}, \ and\ \bibinfo
  {author} {\bibfnamefont {J.~P.}\ \bibnamefont {Perdew}},\ }\href@noop {}
  {\bibfield  {journal} {\bibinfo  {journal} {Phys. Rev. B}\ }\textbf {\bibinfo
  {volume} {85}},\ \bibinfo {pages} {014111} (\bibinfo {year}
  {2012})}\BibitemShut {NoStop}%
\bibitem [{\citenamefont {Santra}\ \emph {et~al.}(2008)\citenamefont {Santra},
  \citenamefont {Michaelides}, \citenamefont {Fuchs}, \citenamefont
  {Tkatchenko}, \citenamefont {Filippi},\ and\ \citenamefont
  {Scheffler}}]{Santra:2008jw}%
  \BibitemOpen
  \bibfield  {author} {\bibinfo {author} {\bibfnamefont {B.}~\bibnamefont
  {Santra}}, \bibinfo {author} {\bibfnamefont {A.}~\bibnamefont {Michaelides}},
  \bibinfo {author} {\bibfnamefont {M.}~\bibnamefont {Fuchs}}, \bibinfo
  {author} {\bibfnamefont {A.}~\bibnamefont {Tkatchenko}}, \bibinfo {author}
  {\bibfnamefont {C.}~\bibnamefont {Filippi}}, \ and\ \bibinfo {author}
  {\bibfnamefont {M.}~\bibnamefont {Scheffler}},\ }\href@noop {} {\bibfield
  {journal} {\bibinfo  {journal} {J. Chem. Phys.}\ }\textbf {\bibinfo {volume}
  {129}},\ \bibinfo {pages} {194111} (\bibinfo {year} {2008})}\BibitemShut
  {NoStop}%
\bibitem [{\citenamefont {Benedict}\ \emph {et~al.}(1956)\citenamefont
  {Benedict}, \citenamefont {Gailar},\ and\ \citenamefont
  {Plyler}}]{Benedict:1956id}%
  \BibitemOpen
  \bibfield  {author} {\bibinfo {author} {\bibfnamefont {W.~S.}\ \bibnamefont
  {Benedict}}, \bibinfo {author} {\bibfnamefont {N.}~\bibnamefont {Gailar}}, \
  and\ \bibinfo {author} {\bibfnamefont {E.~K.}\ \bibnamefont {Plyler}},\
  }\href@noop {} {\bibfield  {journal} {\bibinfo  {journal} {J. Chem. Phys.}\
  }\textbf {\bibinfo {volume} {24}},\ \bibinfo {pages} {1139} (\bibinfo {year}
  {1956})}\BibitemShut {NoStop}%
\bibitem [{\citenamefont {Dyke}\ and\ \citenamefont
  {Muenter}(1973)}]{Dyke:1973bv}%
  \BibitemOpen
  \bibfield  {author} {\bibinfo {author} {\bibfnamefont {T.~R.}\ \bibnamefont
  {Dyke}}\ and\ \bibinfo {author} {\bibfnamefont {J.~S.}\ \bibnamefont
  {Muenter}},\ }\href@noop {} {\bibfield  {journal} {\bibinfo  {journal} {J.
  Chem. Phys.}\ }\textbf {\bibinfo {volume} {59}},\ \bibinfo {pages} {3125}
  (\bibinfo {year} {1973})}\BibitemShut {NoStop}%
\end{thebibliography}%

\end{document}